\documentclass[utf8]{aa}
\usepackage{graphicx}
\usepackage[varg]{txfonts}
\usepackage{natbib}
\bibpunct{(}{)}{;}{a}{}{,}
\newcommand{\ton}{Ton\,345\xspace}
\newcommand{\catwo}{\ion{Ca}{ii}\xspace}
\newcommand{\irt}{Ca\,IRT\xspace}
\newcommand{\acdc}{\textsc{AcDc}\xspace}
\DeclareMathOperator{\total}{d\!}

\begin{document}

\title{Non-LTE spectral models for the gaseous debris-disk component of \ton\thanks{Based on observations collected at the Centro Astronómico Hispano Alemán (CAHA) at Calar Alto, operated jointly by the Max-Planck-Institut für Astronomie and the Instituto de Astrofísica de Andalucía (CSIC).}}

\author{S.~Hartmann
  \and
  T.~Nagel
  \and
  T.~Rauch
  \and
  K.~Werner
}

\institute{Institute for Astronomy and Astrophysics,
  Kepler Center for Astro and Particle Physics,
  University of Tübingen,\\
  Sand~1,
  72076~Tübingen,
  Germany,\\
  \email{hartmann@astro.uni-tuebingen.de}
}

\date{Received 21 February 2014 / accepted 30 September 2014}


\abstract
    {For a fraction of single white dwarfs with debris disks, an additional gaseous disk was discovered. Both dust and gas are thought to be created by the disruption of planetary bodies.}
    {The composition of the extrasolar planetary material can directly be analyzed in the gaseous disk component, and the disk dynamics might be accessible by investigating the temporal behavior of the \catwo infrared emission triplet, hallmark of the gas disk.}
    {We obtained new optical spectra for the first helium-dominated white dwarf for which a gas disk was discovered~(\object{Ton 345}) and modeled the non-LTE spectra of viscous gas disks composed of carbon, oxygen, magnesium, silicon, sulfur, and calcium with chemical abundances typical for solar system asteroids. Iron and its possible line-blanketing effects on the model structure and spectral energy distribution was still neglected. A set of models with different radii, effective temperatures, and surface densities as well as chondritic and bulk-Earth abundances was computed and compared with the observed line profiles of the \catwo infrared triplet.}
    {Our models suggest that the \catwo emission stems from a rather narrow gas ring with a radial extent of $R=0.44\text{--}0.94\,R_{\sun}$, a uniform surface density $\varSigma=0.3\,\mathrm{g}\,\mathrm{cm}^{-2}$, and an effective temperature of $T_{\mathrm{eff}}\approx 6000\,\mathrm{K}$. The often assumed chemical mixtures derived from photospheric abundances in polluted white dwarfs -- similar to a chondritic or bulk-Earth composition -- produce unobserved emission lines in the model and therefore have to be altered. We do not detect any line-profile variability on timescales of hours, but we confirm the long-term trend over the past decade for the red-blue asymmetry of the double-peaked lines.}
    {}


    \maketitle
    %
    \section{Introduction}
    When the effective temperature~($T_{\mathrm{eff}}$) of a white dwarf~(WD) decreases, its atmosphere should consist almost purely of either hydrogen or helium \citep[e.g.,][]{Koester:2009}. The efficiency of gravitational settling causes any heavier element to sink below the photospheric layer, where it is no longer observable. The $20\text{--}30\mathrm{\%}$ of single WDs \citep{Zuckerman:2010} with $T_{\mathrm{eff,\,WD}}\lesssim 25\,000\,\mathrm{K}$ that show metallic pollution must therefore actively accrete matter from their vicinity. Shortly after they were discovered, the interstellar medium was the assumed source for the accreted metals \citep{Dupuis:1992}, since then the explanation has shifted considerably. Today, there is coherent evidence that the WDs host dusty disks within their stellar tidal radius \citep{Zuckerman:1987,Becklin:2005,Kilic:2005,Kilic:2006,Farihi:2007,Farihi:2009,Farihi:2010}, which provide the repository for continuous accretion process.

    Spectroscopic analyses of the WD photospheric absorption lines produced by the accreted material provide further indications about the origin of the dust. Its chemical composition seems to be similar to that of rocky material in the inner solar system \citep{Zuckerman:2007,Klein:2010,Dufour:2010,Farihi:2011}. Therefore, the favored scenario proposed for the creation of the disks is that a smaller planetary body like an asteroid is scattered toward the central object and destroyed by the tidal forces \citep{Debes:2002,Jura:2003}. Although the first results gained by photospheric measurements are impressive and provide an inside view into extrasolar planetary systems in the post-main sequence phase of a host star, there is a drawback to this method. The metal-diffusion rates in the WD atmosphere are difficult to obtain \citep{Koester:2009} and other possible effects have to be considered. Thermohaline convection can affect the abundance analysis for hydrogen-rich (DA-type) WDs \citep{Deal:2013}, and radiative levitation can become important at low mass-accretion rates \citep{Chayer:2014}.

    In the past years, gaseous metal-rich disks orbiting hydrogen-rich \citep[e.g., \object{SDSS J122859.93+104033.0},][hereafter SDSS\,J1228]{Gaensicke:2006} and helium-rich \citep[e.g., \ton,][]{Gaensicke:2008} single WDs were reported. These disks can be recognized by the typically double-peaked, infrared emission \catwo triplet~(\irt) at $\lambda\lambda\ 8498$, $8542$, and $8662\,\text{\AA}$. The explanation of the disk origin is a variation of the scenario described above. The dusty planetary debris particles collide with each other with high relative velocity \citep{Jura:2008}, producing the gas content. Alternatively, \citet{Rafikov:2011} employed a model in which gas is produced via sublimating the material at the dust disk inner radius. In both cases viscous torques spread the gas in- and outward, forming a gaseous secondary disk. And indeed, all known gas debris disks around single WDs have been found to spatially match a corresponding dusty disk \citep{Melis:2010,Brinkworth:2012}. Because the gas naturally has a higher effective temperature of about $T_{\mathrm{eff}}\approx 5000\,\mathrm{K}$, the resulting spectral features yield an opportunity to analyze the debris material directly and before it is accreted onto the WD atmosphere.

    In our analyses of the gaseous disk component around the hydrogen-rich WD SDSS\,J1228 \citep{Werner:2009,Hartmann:2010,Hartmann:2011}, we especially investigated the reported variations \citep{Gaensicke:2006} of the hallmarking \irt spectral feature, which might yield information about the disk geometry and temporal development. Similar time variations were noted for \ton \citep{Gaensicke:2008}, the first helium-dominated WD for which a gaseous debris disk was reported. Consequently, we applied for new optical Calar~Alto~Observatory observations of \ton to investigate the dynamical processes of the gaseous disk component. We used state-of-the-art, non-local thermodynamic equilibrium~(NLTE) models and synthetic spectra calculated with the accretion disk code \acdc \citep{Nagel:2004} to determine the influence of input parameters on the \irt and compared its theoretical line profiles with our observations.

    In the following, we introduce the observed object (Sect.\,\ref{sec:ton345}) and briefly describe the observations and the data reduction (Sect.\,\ref{sec:obs}). In Sect.\,\ref{sec:var}, we examine the short- and long-time variability of the obtained spectra. The method we used to model the disk is introduced in Sect.\,\ref{sec:acdc}. We then (Sect.\,\ref{sec:result}) investigate the influence of various model parameters, such as the chemical mixture and the effective temperature, on the calculated synthetic spectra of a gaseous debris disk and compare our best-fit model with the observations. Finally, we conclude in Sect.\,\ref{sec:conclusion} with an outlook on necessary future work.

    \section{\ton}\label{sec:ton345}
    \ton (\object{WD 0842+231}, \object{SDSS J084539.17+225728.0}) was discovered by \citet{Iriarte:1957} in their Tonantzintla Observatory blue star survey. Later, it was also included in the Palomar-Green survey \citep{Green:1986} and the hot-subdwarf catalog \citep{Kilkenny:1988}. In their automatic search for circumstellar gaseous disks around $15\,000$ blue ($i-g<0.8$) stellar objects taken from the sixth data release of the Sloan Digital Sky Survey \citep[SDSS,][]{Adelman:2008} \citet{Gaensicke:2008} found \ton to exhibit a significant ($3\sigma$) excess of the \irt. Because no evidence of a stellar companion was found, they concluded that \ton was one of the first identified single WDs hosting a gaseous metal-rich disk; it also ist the first helium-dominated WD (spectral type DBZ).

    From analyzing the chemical composition of the WD, \citet{Gaensicke:2008} gave an upper limit for the hydrogen content of $\text{H/He}<3\times 10^{-5}$ (by number) and abundance values for calcium ($\text{Ca/He}=1.3\times 10^{-7}$), magnesium $(\text{Mg/He}=6.0\times 10^{-6})$, and silicon $(\text{Si/He}=8.0\times 10^{-6})$. The best model fit was achieved for $T_{\mathrm{eff,\,WD}}=18\,500\,\mathrm{K}$ and $\log{g}=8.3$, corresponding to a mass of $M_{\mathrm{WD}}=0.70\,M_{\sun}$.
    The gas disk \irt is significantly weaker than for SDSS\,J1228, but the asymmetry of the Doppler-shifted blue and red line-peaks is more prominent for Ton\,345. Additionally, \citet{Gaensicke:2008} found the \ion{Fe}{ii} $\lambda\ 5169\,\text{\AA}$ emission line that is also observed in SDSS\,J1228, but no other spectral signatures of the gaseous disk have been detected. According to their modeling method \citep{Gaensicke:2006}, the best fit was achieved for a $\varDelta R_{\mathrm{disk}}=0.5\text{--}1.0\,R_{\sun}$ wide, eccentric $(\epsilon=0.2\text{--}0.4)$ disk structure. \citet{Melis:2010} derived similar values for the gas disk radius and temperature from their Keck HIRES\footnote{High resolution echelle spectrometer.} observations.

    \section{Observation and data reduction}\label{sec:obs}
    Our observations of \ton were performed on UTC\,2011-04-08 with the Cassegrain Twin Spectrograph~(TWIN) at the $3.5\,\mathrm{m}$ telescope at \mbox{Calar~Alto~Observatory}. We used the gratings T08 for the blue and T04 for the red channel, both with a dispersion of $72\,\text{\AA}\,\mathrm{mm}^{-1}$ (resolution of about $2.5\,\text{\AA}$). The resulting spectra cover the wavelength ranges $3500\text{--}6500\,\text{\AA}$ and $5500\text{--}9000\,\text{\AA}$. Seventeen consecutive spectra were obtained simultaneously for both ranges, with $900\,\mathrm{s}$ integration time each.

\begin{figure}
      \centering
      \resizebox{\hsize}{!}{\includegraphics{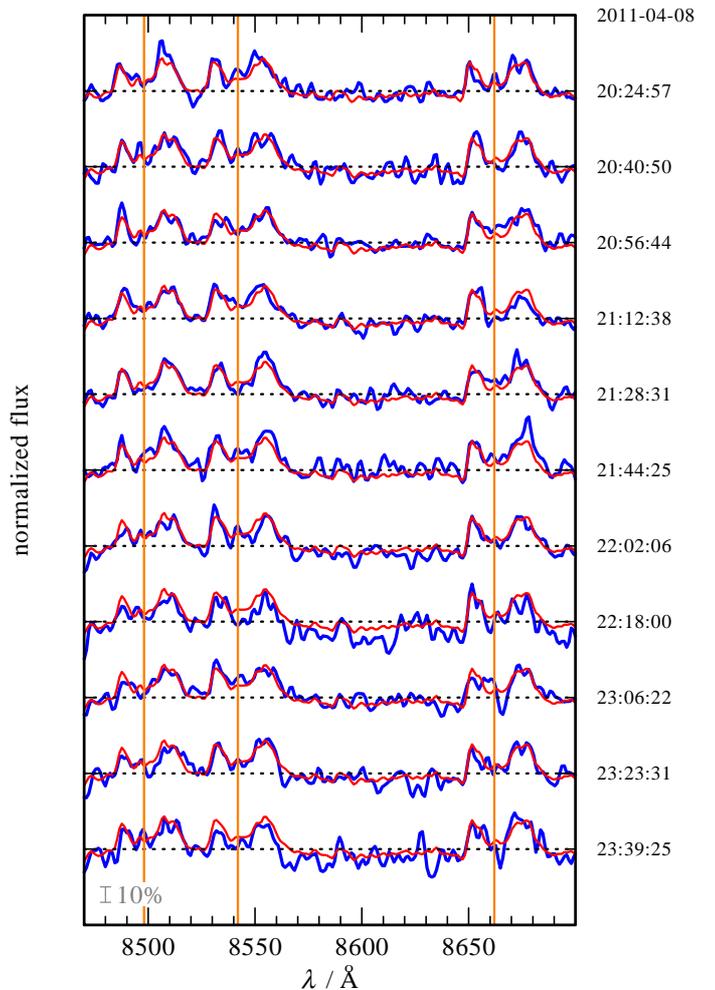}}
      \caption{\catwo infrared triplet in the 900\,s-exposure normalized spectra of \ton (blue in the online version, thick line), overlaid by the co-added spectrum (red, thin line). Vertical lines (orange) indicate the rest wavelengths of the triplet. The dotted, horizontal lines display the normalized flux value of unity for each observation. The vertical bar represents $10\mathrm{\%}$ of the continuum flux.}
      \label{fig:short}
    \end{figure}

    The exposures were reduced and extracted using IRAF\footnote{IRAF is distributed by the National Optical Astronomy Observatory, which is operated by the Association of Universities for Research in Astronomy (AURA) under cooperative agreement with the National Science Foundation.} standard tasks and the available extinction and standard star data. The continuously rising air mass $(1.04\text{--}2.25)$ during the night caused the quality of the spectra to deteriorate, respectively. Therefore, only eleven exposures with a signal-to-noise ratio of $\text{S/N}>6$, shown in Fig.\,\ref{fig:short}, were considered in our analysis and were co-added to increase the ratio.

    \begin{figure*}
      \centering
      \resizebox{\hsize}{!}{\includegraphics[angle=-90]{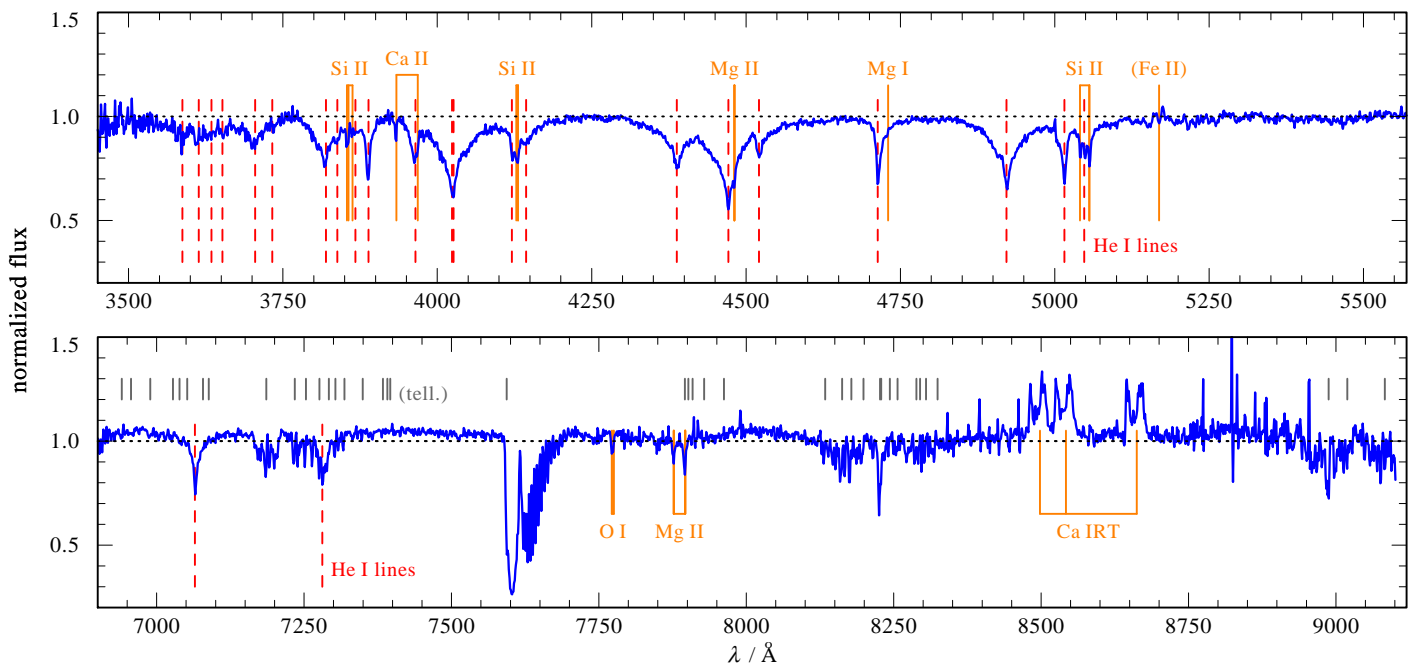}}
      \caption{Normalized, co-added spectrum of Ton\,345. The (red) dashed, vertical lines show the photospheric \ion{He}{i} absorption lines. Other absorption features from the metal pollution of the WD atmosphere and the disk \irt emission are indicated by (orange) solid, vertical lines. In the lower panel the telluric lines are marked by the (gray) upper tickmarks.}
      \label{fig:total}
    \end{figure*}

    Typical for a DB(Z) WD, the blue ($3500\text{--}6500\,\text{\AA}$, upper panel of Fig.\,\ref{fig:total}) part of the combined spectrum is dominated by \ion{He}{i} lines. The red part of the observation is intermingled with the strong telluric lines of $\text{H}_{2}\text{O}$ and $\text{O}_{2}$. Additional absorption features result from the pollution of the WD atmosphere by O, Si, Mg, and Ca. The gaseous disk is only indicated by the prominent \irt lines near $8600\,\text{\AA}$. The previously reported \ion{Fe}{ii} emission line at $5169\,\text{\AA}$ cannot be seen unambiguously in our data.

    \section{Line-profile variability}\label{sec:var}
\begin{figure}
      \centering
      \resizebox{\hsize}{!}{\includegraphics{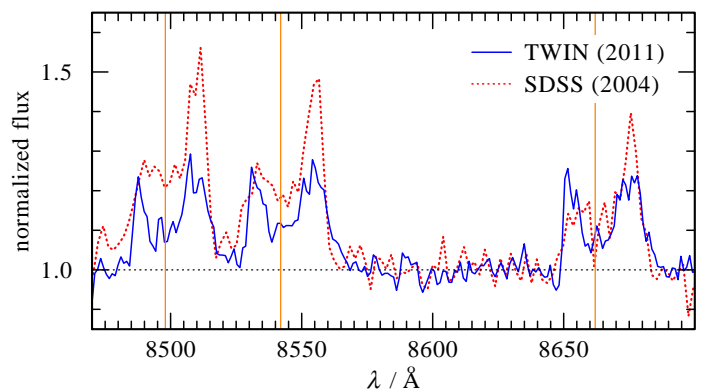}}
      \caption{Comparison of the normalized, co-added best TWIN exposures~(blue, solid line) and the SDSS data from 2004~(red, dotted line) in the spectral range of the \irt (rest wavelength marked by the orange, vertical lines).}
      \label{fig:coadd}
    \end{figure}

    Like SDSS\,J1228 \citep{Gaensicke:2006}, the \irt of \ton changed over time. When comparing the SDSS observation taken in 2004 December (Fig.\,\ref{fig:coadd}) with William Herschel Telescope spectra taken in 2008 January, \citet{Gaensicke:2008} noted a reduction in the line asymmetry and equivalent width. Although the redshifted parts remain the prominent feature of all three lines, both peaks are clearly recognizable and of a more similar maximum height in the later observations. The Keck HIRES data taken in the same year confirmed the changes \citep{Melis:2010}.

    In our co-added spectrum (Fig.\,\ref{fig:coadd}), the reported change compared with the data from 2004 is clearly visible. The asymmetric peak-height of both components vanished even more than in the spectra of \citet{Gaensicke:2008} and \citet{Melis:2010}.
    Furthermore, the peak widths of each line have altered significantly. Fitting the line components with two simple Lorentzians, we determined the full width at half maximum (FWHM) of each of the three red and blue components in the SDSS and Calar Alto observations. The mean value for the red peaks increased from $8.1\pm 2.1\,\text{\AA}$ in the 2004 data to $10.7\pm 2.3\,\text{\AA}$ in the latter. In contrast, the average FWHM of the blue components dropped from $14.5\pm 5.3\,\text{\AA}$, with almost no recognizable peak in 2004, to $5.7\pm 1.2\,\text{\AA}$ in the Calar Alto data of 2011. The explanation of this long-term change in peak-height and -width is still lacking.

    The (gaseous) debris disks around single WDs are thought to be formed by the disruption of planetesimal bodies passing their host star on close, eccentric orbits. Therefore, the resulting disk is not be expected to be of axisymmetric shape. During the disruption phase, some parts of the originally solid object may have been scattered onto wider orbits than others. The material experiences different strengths of the central gravitational force. Hydrodynamic models \citep{Hartmann:2010,Hartmann:2011} suggest the formation of spiral-arm or off-center geometry.

    An asymmetric distribution of the disk mass produces asymmetric double-peaked spectral lines as radiation from differently sized areas of material becomes red- and blueshifted. The rotation of such a disk would result in changes in the line profiles on Keplerian timescales. For the disk of \ton, which orbits roughly at $R_{\sun}$, this time should be of the order of hours and recognizable within our observation night (Fig.\,\ref{fig:short}). Comparing the single exposures with the co-added spectrum, however, there is no sign for such a short-term variability, although the poor S/N ratio may mask the effect.

    \section{Accretion disk code}\label{sec:acdc}
    To model the spectrum of the gaseous debris disk component of \ton, we used our accretion disk code \acdc \citep{Nagel:2004}. In a first step, we assumed an axially symmetric, geometrically thin accretion disk. Such a disk can be separated into concentric rings of plane-parallel geometry, that can be handled independently of each other. Therefore, the radiative transfer becomes a one-dimensional problem. Each ring is described by its radius $R$, the effective temperature $T_{\mathrm{eff}}(R)$, the surface mass density $\varSigma(R)$, and the chemical composition. \acdc is designed to assume the emitted radiation to be viscously generated, so the Reynolds number ${Re}(R)$ (or the viscosity parameter $\alpha(R)$) enters as an additional parameter. For all models presented here, we assumed ${Re}=15\,000$. For such a viscous $\alpha$-disk, the radial run of $T_{\mathrm{eff}}(R)$ and $\varSigma(R)$ can be expressed in terms of the mass-accretion rate $\dot{M}(R)$ and mass $M_{\mathrm{WD}}$ and radius $R_{\mathrm{WD}}$ of the central star \citep{Shakura:1973}.

    For each of the disk rings, radiation transport and vertical structure are described by the following set of equations:
    \begin{itemize}
    \item The radiation transfer for the specific intensity $I$ at frequency $\nu$ is calculated along a ray with inclination angle $\theta$ between the ray and the normal vector of the disk midplane
      \begin{equation}
        \mu\,\frac{\partial\,I(\nu,\mu,z)}{\partial\,z}=-\chi(\nu,z)I(\nu,\mu,z)+\eta(\nu,z)\quad{,}
      \end{equation}
      with $z$ being the height above the midplane. Here, $\chi$ is the absorption coefficient, $\eta$ the emission coefficient, and $\mu=\cos{\theta}$.
    \item The hydrostatic equilibrium of gravitation, gas pressure $P_{\mathrm{gas}}$, and radiation pressure is given by
      \begin{equation}
        \frac{\total P_{\mathrm{gas}}}{\total m}=\frac{GM_{\mathrm{WD}}}{R^{3}}\,z-\frac{4\piup}{c}\!\int\limits_{0}^{\infty}\frac{\chi(\nu)}{\rho}H(\nu,z)\total\nu\quad{,}
      \end{equation}
      with $\rho$ denoting the mass density and $H$ the Eddington flux. We introduced the column-mass density $m$ as
      \begin{equation}
        m(z)=\int\limits_{z}^{\infty}\rho(z^{\prime})\total z^{\prime}\quad{.}
      \end{equation}
    \item The energy balance between the radiative energy loss $E_{\mathrm{rad}}$ and the viscously generated energy $E_{\mathrm{mech}}$ is
      \begin{equation}
        E_{\mathrm{mech}}=E_{\mathrm{rad}}
      \end{equation}
      with
      \begin{equation}
        E_{\mathrm{rad}}=4\piup\!\int\limits_{0}^{\infty}\Big(-\chi\left(\nu,z\right)J(\nu,z)+\eta\left(\nu,z\right)\Big)\total\nu
      \end{equation}
      and
      \begin{equation}
        E_{\mathrm{mech}}=w\varSigma\left(R\frac{\total \omega}{\total R}\right)^{2}=\frac{9}{4}\frac{\varSigma\sqrt{R}}{Re}\left(\frac{\sqrt{GM_{\mathrm{WD}}}}{R}\right)^{3}
      \end{equation}
      with the mean intensity $J$, angular velocity $\omega$, and $w$ the kinematic viscosity written following \citet{Lynden:1974}.
    \item The NLTE rate equations for the population numbers $n_{i}$ of the atomic levels $i$ are
      \begin{equation}
        n_{i}\sum_{i\neq j}P_{ij}\,-\,\sum_{j\neq i}n_{j}\,P_{ji}=0\quad{,}
      \end{equation}
      where $P_{ij}$ denotes the rate coefficients, consisting of radiative and electron collisional components.
    \end{itemize}

    Because these equations are highly coupled, \acdc solves them simultaneously in an iterative scheme under the constraints of particle number and charge conservation using the accelerated lambda iteration formalism \citep[e.g.,][]{Werner:1985}.

    To calculate the spectrum of the whole disk, the ring intensities are integrated with respect to the inclined, emitting surface. For a disk ranging from $R_{\mathrm{min}}$ as inner radius to $R_{\mathrm{max}}$ as the outer radius and an inclination angle $i$ this is given by
    \begin{equation}
      I_{\mathrm{disk}}(\nu,i)=\cos i\int\limits_{R_{\mathrm{min}}}^{R_{\mathrm{max}}}\int\limits_{0}^{2\piup}I_{\mathrm{ring}}(\nu,\phi,R)\,R\total\phi\total r\quad{,}
    \end{equation}
    where $\phi$ is the azimuthal angle.

    Befor this, the emergent intensity of each ring given by $I(\nu,\mu=\cos i,z_{\mathrm{max}})$ has to be Doppler shifted to account for the assumed Keplerian motion of the emitting material:
    \begin{equation}
      I(\nu)\mapsto I(\nu^{\prime})\quad\text{with}\quad\nu^{\prime}=\nu\sqrt{\frac{1-v_{i}}{1+v_{i}}}\quad{,}
    \end{equation}
    where $v_{i}$ is the radial velocity $v_{r}$ corrected for the disk inclination $i$:
    \begin{equation}
      v_{i}=v_{r}\sin i=\sqrt{\frac{GM_{\mathrm{WD}}}{R}}\cos\phi \sin i\quad{.}
    \end{equation}

    To reproduce the rather weak observed \irt emission of \ton, all of our models presented in this work have an inclination of $i=87^{\circ}$. A similar result for the dust disk inclination angle with $i=70\text{--}85^{\circ}$ was reported by \cite{Brinkworth:2012}.

    \section{Results}\label{sec:result}
    To study the disk parameters, we calculated models for a chondritic as well as a bulk-Earth-like chemical mixture. Both compositions are well motivated by the photospheric measurements \citep[e.g.,][]{Jura:2006,Jura:2014} of planetary-material-accreting WDs. Iron is not yet included. The abundances used in our models are based on \citet{Lodders:2003} for the chondritic and \citet{McDonough:2003} in the bulk-Earth case. The values, normalized to unity, are given in Table\,\ref{tab:abund}. H and C (for bulk Earth) are only included to use the same atomic data set for both compositions. All atomic data (summarized in Table\,\ref{tab:adata}) are taken from the Tübingen Model Atom Database\footnote{TMAD, \url{http://astro.uni-tuebingen.de/~TMAD}}.

    \begin{table}
      \caption{Abundances of all elements included in the disk models.}
      \label{tab:abund}
      \centering
      \begin{tabular}{lr@{.}lr@{.}l}\hline\hline
        &\multicolumn{4}{c}{$\mathrm{\%}$ mass fraction}\rule{0ex}{2.2ex}           \\\cline{2-5}
        element&\multicolumn{2}{l}{chondrite}&\multicolumn{2}{l}{bulk Earth}\rule{0ex}{2.2ex}   \\\hline
        H&          0&$13\times 10^{-5}$&          0&$14\times 10^{-5}$\rule{0ex}{2.5ex}\\
        C&          4&70               &          0&$14\times 10^{-5}$                 \\
        O&         60&15               &         47&78                                \\
        Mg&        12&62               &         23&33                                \\
        Si&        14&11               &         25&56                                \\
        S&          7&18               &          0&83                                \\
        Ca&         1&24               &          2&50                                \\\hline
      \end{tabular}
    \end{table}

    \begin{table}
      \caption{Overview of the ionization stages, levels, and line transitions used in the atomic data set of our disk models.}
      \label{tab:adata}
      \centering
      \begin{tabular}{l@{\,}lrrr}\hline\hline
        \multicolumn{2}{l}{ions}& NLTE levels& LTE levels& line transitions\rule{0ex}{2.2ex}\\\hline
        H&  \textsc{i--ii} &          11&          6&               45\rule{0ex}{2.2ex}\\
        C&  \textsc{i--iii}&          33&         48&               51\\
        O&  \textsc{i--iii}&          31&        113&               39\\
        Mg& \textsc{i--iii}&          33&         37&               60\\
        Si& \textsc{i--iv} &          57&         33&               91\\
        S&  \textsc{i--iii}&          31&          0&               87\\
        Ca& \textsc{i--iii}&          22&         51&               24\\\hline
      \end{tabular}
    \end{table}

    The models include up to 25 rings, equidistantly distributed between $R_{\mathrm{min}}=0.17\,R_{\sun}$ and $R_{\mathrm{max}}=1.49\,R_{\sun}$. We assumed the disk to be non-stationary, but with constant $\varSigma(R)$ and a flat temperature profile with $T_{\mathrm{eff}}(R)$ (variations smaller than $10\mathrm{\%}$).

    \subsection{Surface density and effective temperature}\label{ssec:temperature}
\begin{figure}
      \centering
      \resizebox{\hsize}{!}{\includegraphics{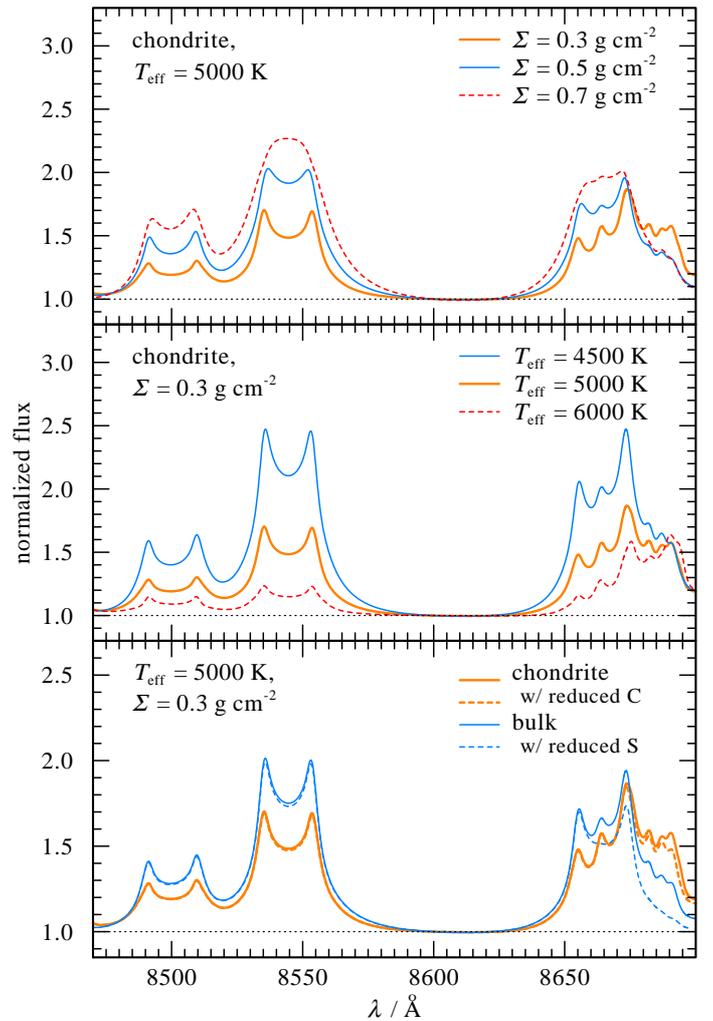}}
      \caption{Normalized synthetic spectra in the range of the \irt for different $\varSigma$ (top), $T_{\mathrm{eff}}$ (middle), and chemical mixtures (bottom). The unmodified abundance values are given in Table\,\ref{tab:abund}.}
      \label{fig:para}
    \end{figure}

    As noted in Sect.\,\ref{sec:acdc}, $\varSigma$ is a free parameter in the model calculations. The top panel of Fig.\,\ref{fig:para} shows three synthetic spectra with different $\varSigma$. We chose the chondritic values of Table\,\ref{tab:abund} for the chemical composition and $T_{\mathrm{eff}}(R)\approx 5000\,\mathrm{K}$. While higher $\varSigma$ tend to increase the \catwo emission in general, they also broaden the lines. This results in a diminishing peak separation when applying the Doppler shift. For the highest value of $\varSigma$ the depression between the peaks mostly vanishes for the $8542$ and $8662\,\text{\AA}$ components. First attempts to use $\varSigma<0.3\,\mathrm{g}\,\mathrm{cm}^{-2}$ for even deeper separation gaps resulted in numerical instabilities. We also found a strong blend of the $8662\,\text{\AA}$ \irt component with adjacent \ion{C}{ii} doublet ($\lambda\lambda\ 8683$, $8697\,\text{\AA}$)  and \ion{S}{i} multiplet ($\lambda\lambda\ 8670$, $8671$, $8679$, $8680$, $8693$, $8694$, $8695\,\text{\AA}$) emission lines, which we address below in Sect.\,\ref{ssec:abund}.

    Furthermore, $T_{\mathrm{eff}}$ is another free parameter in our disk description. Again, we modeled disks with different, but radially almost constant values of $T_{\mathrm{eff}}(R)$. The spectra presented in the center panel of Fig.\,\ref{fig:para} were calculated with $T_{\mathrm{eff}}(R)\approx 4500\,\mathrm{K}$, $5000\,\mathrm{K}$, and $6000\,\mathrm{K}$. All three models are of chondritic chemical mixture according to Table\,\ref{tab:abund} and have the same constant surface density of $\varSigma(R)=0.3\,\mathrm{g}\,\mathrm{cm}^{-2}$. The equivalent width of the \irt decreases rapidly with rising temperature. All three components also have more similar line strengths for models with higher $T_{\mathrm{eff}}$. For the hottest models the previously mentioned blending \ion{C}{ii} and \ion{S}{i} lines become even stronger than the \catwo triplet.

    \subsection{Metal-rich disk composition}\label{ssec:abund}
    In the bottom panel of Fig.\,\ref{fig:para}, the normalized, synthetic flux of models with different abundance patterns (Table\,\ref{tab:abund}) is shown in the range of the \irt. Both the chondritic and the bulk-Earth-like mixture produce fairly similar line profiles, considering width and relative strength of the triplet lines. As noted before, the $8662\,\text{\AA}$ component is distinctly mingled with emission lines of \ion{C}{ii} and \ion{S}{i}.

    Reducing the C abundance in the chondritic case to $4.7\times 10^{-3}$ (mass fraction), which is $10\mathrm{\%}$ of the starting value, has only a minor impact on the blend, which indicates that the C abundance must be significantly lower. In fact, both starting abundances, C and S, have to be scaled down by at least a factor of $10^{-3}$ ($4.7\times 10^{-5}$ and $7.2\times 10^{-5}$ mass fraction) to suppress the blending line features.

    As our bulk-Earth mixture has essentially no C to begin with (for numerical reasons it was set to $1.4\times 10^{-8}$ in mass fraction) and, thus, does not exhibit C emission lines, we reduced S to $10\mathrm{\%}$ and $1\mathrm{\%}$ of the original value and only the $1\mathrm{\%}$ model (i.e., $8.3\times 10^{-5}$, mass fraction) leads to a complete suppression of the blending S emission, hence, this S abundance value is regarded as an upper limit. An upper C abundance limit of $10^{-4}$, mass fraction, was found by increasing the C value in the $1\mathrm{\%}$ S-reduced bulk-Earth model stepwise by one dex until significant \ion{C}{ii} emission appears that would be incompatible with the observation. For this, the back-reaction of the increased C onto the model structure was neglected because C is a trace element compared to the main constituents (O, Mg, Si, Ca).

    \subsection{Model disk radial extent}\label{ssec:geomety}
    Since the double-peak structure of the \irt is due to the disk rotation, the radial extent of the emitting region can be investigated by altering the inner and outer radius of the model disks. We recall, however, that the peak separation is not equivalent to the disk innermost orbit but is instead dominated by the Doppler-shifted signal of the ring with the strongest emission. Instead, the wings of the line profiles are a better indicator for the inner radius of the emission region.

\begin{figure}
      \centering
      \resizebox{\hsize}{!}{\includegraphics{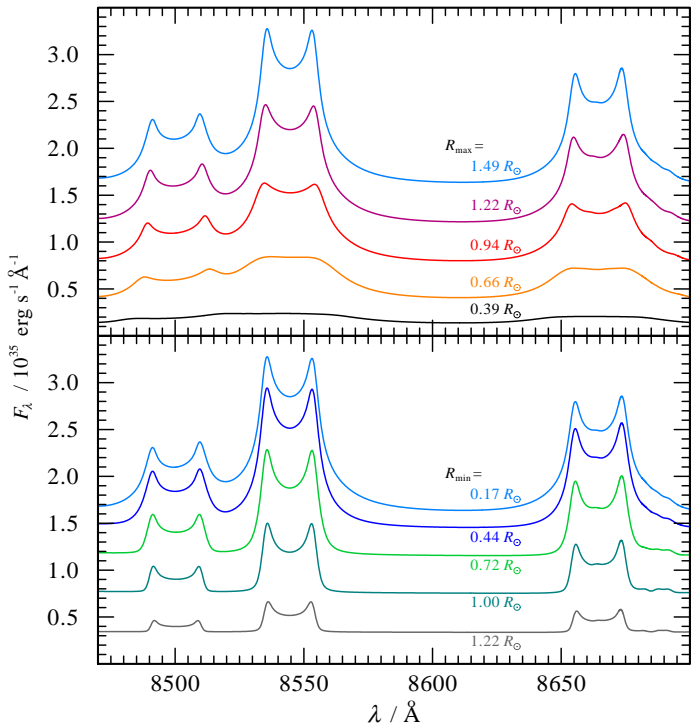}}
      \caption{Spectra of disk models with various outer~(top panel) or inner radius~(bottom panel)\@. In the upper panel, $R_{\mathrm{max}}$ varies while $R_{\mathrm{min}}$ is kept constant at $0.17\,R_{\sun}$. For the models in the lower panel, all with a constant $R_{\mathrm{max}}=1.49\,R_{\sun}$, the inner radius $R_{\mathrm{min}}$ is altered.}
      \label{fig:radial}
    \end{figure}

    Figure\,\ref{fig:radial} shows synthetic spectra of disk models, again with constant $\varSigma(R)=0.3\,\mathrm{g}\,\mathrm{cm}^{-2}$ and $T_{\mathrm{eff}}(R)\approx 5000\,\mathrm{K}$, but with either a varied outer radius $R_{\mathrm{max}}$~(upper panel) or inner radius $R_{\mathrm{min}}$~(lower panel)\@. The respective other radial limit is kept constant. Because of the resulting difference in emitting surface area the continua of all models differ.

    For the models in the upper panel of Fig.\,\ref{fig:radial}, the \irt line-strengths clearly increases with the increase of the disk outer radius. The difference between the \catwo triplet components also grows for larger $R_{\mathrm{max}}$. The separation of the peaks, on the other hand, decreases for the larger disks, suggesting that the strongest emissions originate from the outer regions.

    Shifting the inner radius closer to the central object results in overall stronger \catwo lines and an increase in the relative height difference of the lines to each other. The slopes of the line wings for disks with larger $R_{\mathrm{min}}$ are much steeper, resulting in a flat, continuum-dominated separation between the three components.

    To reduce the influence of the blending metal emission lines, the chemical abundances in the models were set to the S-reduced bulk-Earth-mixture values described in Sect.\,\ref{ssec:abund}. In general, models with the chondritic or the unchanged bulk-Earth composition (not presented here) behave in a similar way with respect to $R_{\mathrm{min}}$ and $R_{\mathrm{max}}$. Additionally, the overlapping metal emission lines redward of $8662\,\text{\AA}$ become more pronounced for disks including more distant rings.

    \subsection{Comparison with observation}\label{ssec:bestfit}
    Combining the parameter studies above, a model disk can be tailored to match the \catwo emission of the observed spectrum. Because of the highly non-linear nature of the NLTE equations to be solved for each ring, many ring models are numerically unstable and often require frequent user intervention to enforce convergence. By reason of this expenditure of time, a multiparameter fit of the disk consisting of several rings has to be limited to simple by-eye validation.

\begin{figure}
      \centering
      \resizebox{\hsize}{!}{\includegraphics{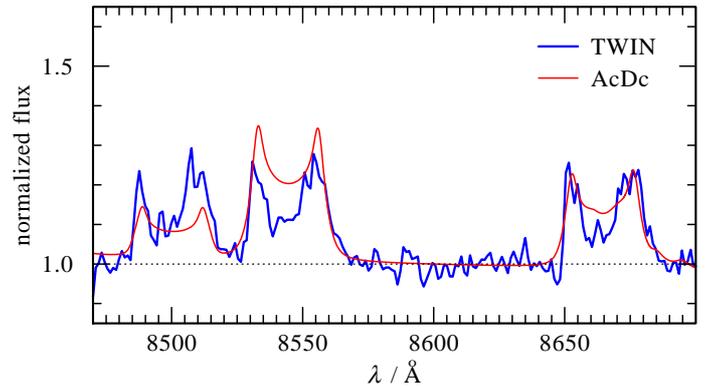}}
      \caption{Comparison of the co-added TWIN data (thick, blue line) and the synthetic spectrum (thin, red line). The model has a radial extent of $R=0.44\text{--}0.94\,R_{\sun}$, constant $\varSigma(R)=0.3\,\mathrm{g}\,\mathrm{cm}^{-2}$, (almost) constant $T_{\mathrm{eff}}(R)=6000\,\mathrm{K}$ and the reduced bulk-Earth chemical mixture.}
      \label{fig:best}
    \end{figure}

    For \ton, we found the synthetic spectrum shown in Fig.\,\ref{fig:best} to sufficiently reproduce our co-added observations of Sect.\,\ref{sec:obs}. The model was calculated for a disk with $R_{\mathrm{min}}=0.44\,R_{\sun}$ and $R_{\mathrm{max}}=0.94\,R_{\sun}$. The assumed radially constant values for the surface density and the effective temperature are $\varSigma(R)=0.3\,\mathrm{g}\,\mathrm{cm}^{-2}$ and $T_{\mathrm{eff}}(R)\approx 6000\,\mathrm{K}$.

    We used the bulk-Earth chemical mixture of Table\,\ref{tab:abund}, but reduced the S abundance to $1\mathrm{\%}$ of the given value. In doing so, the blend of the reddest component of the \irt due to \ion{C}{ii} and \ion{S}{i} lines is minimal. Although the $8662\,\text{\AA}$ component height agrees with the observation, the other two lines show an antipodal behavior. The line at $8498\,\text{\AA}$ is too weak, while the $8542\,\text{\AA}$ emission feature is overestimated.

    \section{Summary and conclusion}\label{sec:conclusion}
    The gaseous component of debris disks around single WDs yields the unique opportunity of directly studing extrasolar planetary material. \ton is the first reported, helium-dominated WD exhibiting the \catwo emission lines at $8600\,\text{\AA}$. Within the past decade, this triplet changed significantly in strength and profile, which may be interpreted as a change in the accretion disk geometry over time.

    In our TWIN observations, we examined the time-frame on which these changes might occur. Short-time changes due to an asymmetric disk rotating around the WD should result in short-time ($\approx$\,hours) variations of the Doppler-shifted double peaks of the \irt. Given the relatively poor signal-to-noise ratio of our spectra, no evidence of line-profile variability is seen. By combining the eleven best exposures we obtained rather symmetric line profiles, which corroborates the reported long-term ($\sim$\,years) trend of a diminishing asymmetry over the past decade. More time-series observations with an $8\,\mathrm{m}$ class telescope are desirable.

    We used \acdc to calculate synthetic spectra for metal-rich, gaseous debris disks and evaluated the influence of several model parameters: the surface density, the effective temperature, the radial extent, and the chemical mixture. We then tailored a model specifically to match the \catwo emission of our co-added TWIN observation. The best by-eye fit was achieved for a geometrically thin, non-stationary $\alpha$-disk with constant $\varSigma(R)=0.3\,\mathrm{g}\,\mathrm{cm}^{-2}$ and $T_{\mathrm{eff}}(R)\approx 6000\,\mathrm{K}$. These model parameters result in \catwo infrared emission lines of similar strength ($T_{\mathrm{eff}}$) and a rather strong depression between the Doppler-shifted peaks of the lines ($\varSigma$).

    The outer radius $R_{\mathrm{max}}=0.94\,R_{\sun}$ of our model is similar to the $R_{\mathrm{max}}\approx 1.0\,R_{\sun}$ found by \cite{Gaensicke:2008}. For the inner disk radius a comparison of our observations and models suggest $R_{\mathrm{min}}=0.44\,R_{\sun}$. The wider line wings of models with a smaller minimal radius, like $R_{\mathrm{min}}=0.27\,R_{\sun}$ reported in \cite{Melis:2010}, would fill up the clearly observed separation between the \irt components significantly.

    From our models computed with bulk-Earth and chondritic composition, we conclude that the actual composition of the gas disk around \ton more resembles a bulk-Earth mixture, which is also favored by indirect analyses of the more common dust-debris disks \citep[e.g.,][]{Jura:2014}. Some modeled metal lines strongly intermingle with the reddest \irt component. Such a blended emission feature is not seen in the observations, suggesting a lower abundance of S\@. A reduction to $1\mathrm{\%}$ of the original value, that is, $8.3\times 10^{-5}$ in mass fraction, suppresses the blending sufficiently. Chondritic models have to be significantly reduced in S and C, and therefore are not as suitable to reproduce the observed data. A more comprehensive study of the possible chemical mixtures -- including iron -- is currently in progress.

    Assuming a circular shape, the total disk surface is $A_{\mathrm{disk}}=\piup (R_{\mathrm{max}}^{2}-R_{\mathrm{min}}^{2})=1.05\times 10^{22}\,\mathrm{cm}^{2}$. With the derived surface density value the total mass of the emitting gas is $M_{\mathrm{disk}}=3.14\times 10^{21}\,\mathrm{g}$. This is similar to masses of well-known, regular planetesimals in our solar system. Considering the additional dust disk coexisting with the gaseous debris disk, this only yields a very much lower limit to the mass of the original body, which was destroyed around \ton.

    \begin{acknowledgements}
      S.H. and T.R. are supported by the DFG (grant We1312/37-1) and the DLR (grant 05\,OR\,1402). This research has made use of the SIMBAD database, operated at the CDS, Strasbourg, France. Funding for SDSS-III has been provided by the Alfred P. Sloan Foundation, the Participating Institutions, the National Science Foundation, and the U.S. Department of Energy Office of Science. The SDSS-III web site is \url{http://www.sdss3.org/}. SDSS-III is managed by the Astrophysical Research Consortium for the Participating Institutions of the SDSS-III Collaboration including the University of Arizona, the Brazilian Participation Group, Brookhaven National Laboratory, Carnegie Mellon University, University of Florida, the French Participation Group, the German Participation Group, Harvard University, the Instituto de Astrofisica de Canarias, the Michigan State/Notre Dame/JINA Participation Group, Johns Hopkins University, Lawrence Berkeley National Laboratory, Max Planck Institute for Astrophysics, Max Planck Institute for Extraterrestrial Physics, New Mexico State University, New York University, Ohio State University, Pennsylvania State University, University of Portsmouth, Princeton University, the Spanish Participation Group, University of Tokyo, University of Utah, Vanderbilt University, University of Virginia, University of Washington, and Yale University.
    \end{acknowledgements}

    \bibliographystyle{aa}
    \bibliography{hartmann}%

\end{document}